\documentclass[%
 reprint,
superscriptaddress,
 amsmath,amssymb,
 aps,
prx,
]{revtex4-2}

\usepackage{graphicx}% Include figure files
\usepackage{dcolumn}% Align table columns on decimal point
\usepackage{bm}% bold math
\usepackage{siunitx}
\DeclareSIUnit\angstrom{\text{\AA}}
\usepackage[hidelinks]{hyperref}% add hypertext capabilities
\usepackage{orcidlink}

\usepackage{ifthen}
\usepackage{amssymb}
\usepackage{xcolor}

\newboolean{showcomments}
\setboolean{showcomments}{true}

\makeatletter
\newcommand{\mynote}[3]{%
  \ifthenelse{\boolean{showcomments}}{%
   \fbox{\bfseries\sffamily\scriptsize#1}%
   {\small$\blacktriangleright$\textsf{\emph{\color{#3}{#2}}}$\blacktriangleleft$}}%
  {%
   \@bsphack
   \@esphack
  }%
}
\makeatother

\definecolor{asparagus}{rgb}{0.53, 0.66, 0.42}

\usepackage{comment}

\begin{document}

\preprint{APS/123-QED}

\title{Origin of the inhomogeneous linewidth in a solid state $^{229}$Th:CaF$_2$ nuclear clock}% Force line breaks with \\

\author{Kai Li\orcidlink{0000-0001-7025-5314}}
\affiliation{%
 JILA, NIST and University of Colorado, Department of Physics,
 University of Colorado, Boulder, Colorado 80309, USA
}

\author{Tian Ooi\orcidlink{0009-0004-2738-0186}}
\affiliation{%
 JILA, NIST and University of Colorado, Department of Physics,
 University of Colorado, Boulder, Colorado 80309, USA
}

\author{Jack F. Doyle\orcidlink{0009-0007-8012-4781}}
\affiliation{%
 JILA, NIST and University of Colorado, Department of Physics,
 University of Colorado, Boulder, Colorado 80309, USA
}

\author{Konstantin Herb\orcidlink{0009-0006-9676-8066}}
\affiliation{%
 JILA, NIST and University of Colorado, Department of Physics,
 University of Colorado, Boulder, Colorado 80309, USA
}

\author{Emil Pellett\orcidlink{0009-0006-1145-8211}}
\affiliation{%
 Department of Physics and Center for Theory of Quantum Matter,
 University of Colorado, Boulder, Colorado 80309, USA
}

\author{Martin Pimon\orcidlink{0000-0001-7784-463X}}
\affiliation{%
 Institute for Theoretical Physics, TU Wien, Stadionallee 2, 1020 Vienna, Austria
}

\author{Thorsten Schumm\orcidlink{0000-0002-1066-202X}}
\affiliation{%
 Institute for Atomic and Subatomic Physics, TU Wien, Stadionallee 2, 1020 Vienna, Austria
}

\author{Steven M. Girvin\orcidlink{0000-0002-6470-5494}}
\affiliation{%
 Departments of Applied Physics and Physics, Yale University,
 New Haven, Connecticut 06511, USA
}
\affiliation{%
 Yale Quantum Institute, Yale University,
 New Haven, Connecticut 06511, USA
}

\author{Leo Radzihovsky\orcidlink{0000-0002-2281-0835}}
\affiliation{%
 Department of Physics and Center for Theory of Quantum Matter,
 University of Colorado, Boulder, Colorado 80309, USA
}

\author{Jun Ye\orcidlink{0000-0003-0076-2112}}
\email{ye@jila.colorado.edu}
\affiliation{%
 JILA, NIST and University of Colorado, Department of Physics,
 University of Colorado, Boulder, Colorado 80309, USA
}


%\affiliation{
%\affiliation{
%\affiliation{%


\date{\today}% It is always \today, today,

\begin{abstract}

Narrow linewidth (high-Q) quantum discriminators are essential for realizing solid-state $^{229}$Th nuclear clocks, yet the observed transition linewidths remain at the tens-to-hundreds-of-kilohertz level, far exceeding the sub-Hertz linewidth set by the radiative lifetime of the nuclear isomer. Here, we investigate the origin of this broadening in $^{229}$Th:CaF$_2$ through systematic measurements of the linewidth as functions of temperature, thorium concentration, and quadrupole state, accompanied by theoretical analysis and first-principle simulations. We find that the linewidth is independent of temperature, increases linearly with thorium concentration, and depends strongly on the quadrupole state. Data analysis identifies variations of the local electric-field gradient (EFG) as the dominant source of inhomogeneous broadening and reveals that the EFG disorder increases linearly with thorium concentration. The observed Lorentzian line shapes and linewidth linear concentration dependence are explained naturally from the $1/r^3$ scaling of a random distribution of long-range dopant-induced perturbations. Complementary atomistic and \textit{ab initio} density-functional-theory calculations reproduce these characteristic features and provide insights into the microscopic defect configurations responsible for the broadening. These results establish the dominant broadening mechanism in solid-state $^{229}$Th and provide guidance for engineering nuclear-clock materials with reduced inhomogeneous broadening.

\end{abstract}

\maketitle

%\tableofcontents

%\section{\label{sec:level1} Introduction}

\section{Introduction} 
Laser spectroscopy has traditionally focused on electronic transitions, whereas nuclear transitions typically occur at photon energies ranging from tens of keV to MeV and are therefore inaccessible to conventional optical laser systems. The $^{229}$Th nucleus is a remarkable exception, featuring a low-lying isomeric state that results from a delicate cancellation of the Coulomb and nuclear contributions to the involved nuclear state energies. The isomeric transition energy is approximately 8.4 eV (148 nm), placing it within the reach of tabletop vacuum-ultraviolet (VUV) laser spectroscopy \cite{kroger1976features,reich1990energy,helmer1994excited,beck2007energy}. In addition, the excited isomeric state possesses a long lifetime of $\sim$$10^3$ s, corresponding to a narrow natural linewidth and high quality factor, making the $^{229}$Th nuclear transition one of the promising candidates for a next-generation optical frequency standard \cite{peik2003nuclear,rellergert2010constraining,kazakov2012performance,derevianko2026colloquium}. Precision spectroscopy of this transition enabled M\"ossbauer-like material studies \cite{zhang2024frequency,zhang2024229thf4,higgins2025temperature,elwell2025laser,ooi2026frequency,hiraki2026laser} and offers a unique platform for probing fundamental physics, including tests of Einstein's equivalence principle, searches for variations of the fine-structure and nuclear interaction coupling constants, as well as searches for dark-matter  \cite{hayes2007sensitivity,peik2021nuclear,fadeev2020sensitivity,fuchs2025searching,arakawa2026probing,Fortier:26}.

Realizing these applications requires precise knowledge of the $^{229}$Th isomer energy and lifetime. Over several decades, extensive efforts have been devoted to finding and accurately determining its nuclear transition energy \cite{helmer1994excited,beck2007energy,von2016direct,thielking2018laser,seiferle2019energy,masuda2019x,yamaguchi2019energy,sikorsky2020measurement}. These studies progressed from indirect spectroscopic measurements to direct resonant laser excitation. The observation of radiative decay established the long-lived nature of the isomeric state \cite{kraemer2023observation,hiraki2024controlling}. Subsequent direct excitation with pulsed VUV lasers localized the transition frequency with an uncertainty at the level of $\sim$10 GHz (limited primarily by the excitation laser linewidth)  in both thorium doped CaF$_2$~\cite{tiedau2024laser} and LiSrAlF$_6$ crystals~\cite{elwell2024laser}, as well as in ThF$_4$ thin films~\cite{zhang2024229thf4}. The high spectral purity VUV frequency-comb immediately improved the transition-frequency determination by more than five orders of magnitude and resolved the quadrupole-split nuclear resonances at a few hundred kilohertz linewidth in the solid-state CaF$_2$ host~\cite{zhang2024frequency}. More recently, continuous-wave VUV absorption spectroscopy has been applied to $^{229}$Th doped CaF$_2$ crystals, achieving similar spectral resolution~\cite{morawetz2026continuous,de2026thorium,huang2026a}.

The solid-state approach embeds a large number of $^{229}$Th nuclei ($\sim$$10^{18}$ cm$^{-3}$) in a VUV-transparent crystal host. Compared against electronic transitions in solids, nuclear transitions have relatively weaker couplings to its environments. Along with the suppression of recoil and first-order Doppler effects in the Lamb-Dicke regime, the nuclear resonance can thus be reasonably narrow within the crystal lattice \cite{rellergert2010constraining,kazakov2012performance}. The ability to interrogate a macroscopic ensemble of nuclei (at the cost of increased linewidth) provides a substantial signal enhancement and offers a promising route toward compact and transportable nuclear clocks. Building on these advances, the frequency stability and temperature dependence of the solid-state $^{229}$Th nuclear clock have been investigated, providing important benchmarks for future clock operation and precision measurements \cite{higgins2025temperature,ooi2026frequency}. In this work, we identify the underlying mechanism that links microscopic configurations to the observed linewidth broadening.

%Two complementary approaches have emerged for realizing a $^{229}$Th nuclear clock: trapped-ion spectroscopy and solid-state spectroscopy. The trapped-ion approach employs established laser-cooling and ion-trapping techniques to isolate individual ions or small ensembles, enabling exceptional control of systematic effects and thus low fractional uncertainty \cite{campbell2009multiply,yamaguchi2024laser}. In contrast, 

\begin{figure*}[t]
\centering
\includegraphics[width=0.9\linewidth]{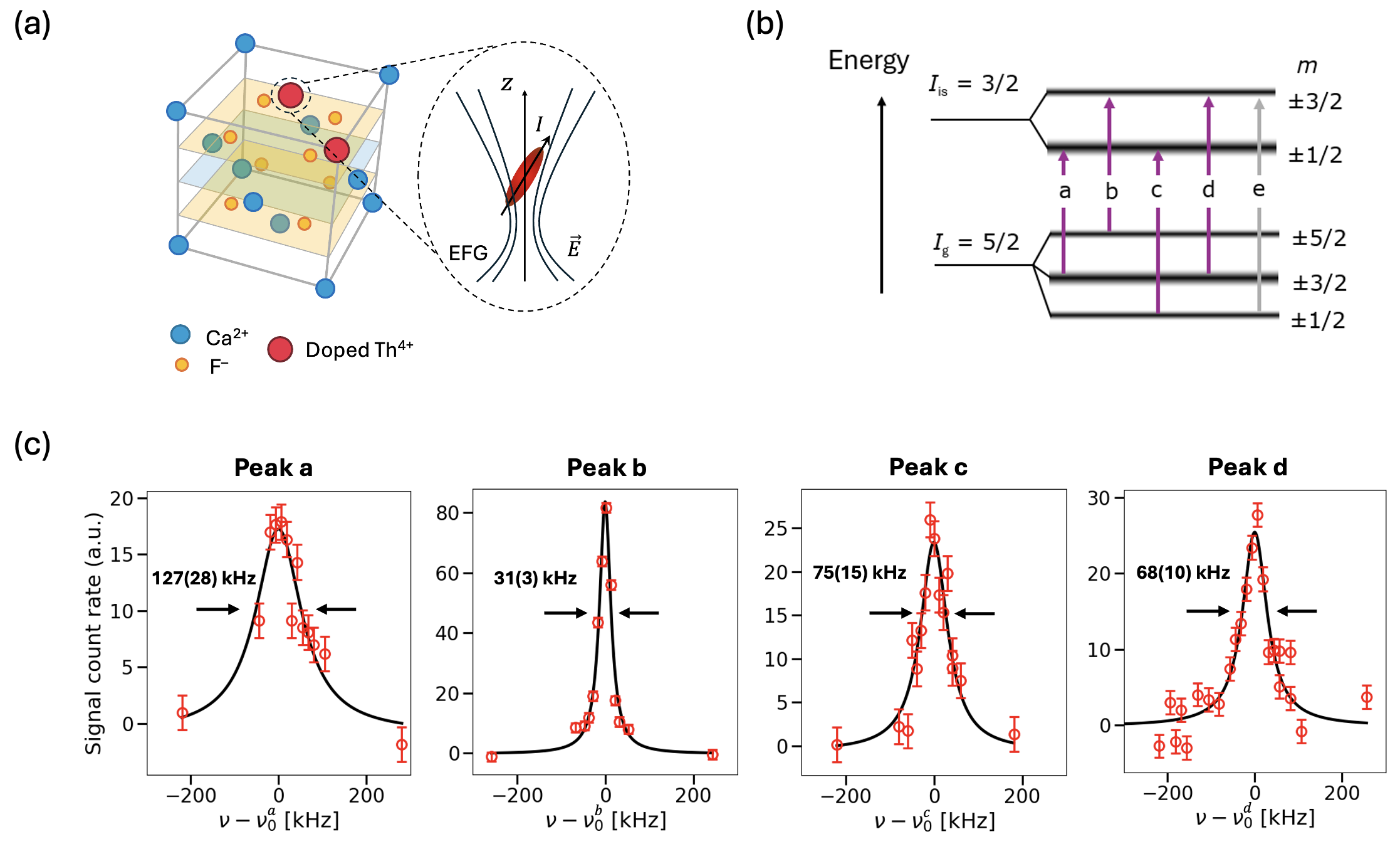}% Here is how to import EPS art
\caption{\label{fig:transitions} 
(a) Schematic of the $^{229}$Th-doped CaF$_2$ crystal. The surrounding charge distribution generates both the electron density at the nucleus and the local EFG, giving rise to the isomeric shift and quadrupole splitting of the nuclear transition, respectively.
(b) The interaction between the nuclear quadrupole moment and the EFG gives rise to quadrupole splitting of the $I_{\mathrm{g}}=5/2$ ground  and $I_{\mathrm{is}}=3/2$ isomeric nuclear states  of $^{229}$Th. Five optical transitions are promoted by the selection rule $\Delta m = 0$ or $\pm 1$, four of which are measured in this study. Transition $e$ has signal around an order of magnitude weaker than the others, due to the relatively small Clebsch–Gordan coefficients, and is not measured in this study.
(c) Representative spectra of the four $^{229}$Th nuclear transitions measured in a CaF$_2$ crystal with a doping concentration of $0.8 \times 10^{18}$ \unit{cm^{-3}} at 195 K. The resonances are surprisingly well described by Lorentzian line shapes (consistent with strain-broadening by point defects~\cite{Stoneham_1969} and also see Appendix~\ref{app:nearest_tail}), with FWHM ranging from tens to hundreds of kHz and exhibiting a strong transition dependence.
}
\end{figure*}

An important step toward improving the performance of solid-state $^{229}$Th nuclear clocks is understanding the mechanisms responsible for broadening the nuclear resonance well beyond its vacuum values. Previous theoretical studies predicted homogeneous broadening on the order of 1 kHz, substantially larger than the millihertz-level natural linewidth, limited primarily by magnetic dipole interactions with F nuclei in the crystal and the second-order Doppler effect at room temperature~\cite{rellergert2010constraining,kazakov2012performance}. Here, we systematically characterize the linewidths of $^{229}$Th nuclear transitions in CaF$_2$ as functions of thorium concentration ($n$), crystal temperature ($T$), and quadrupole state. We observe linewidths ranging from tens to hundreds of kilohertz, 1-2 orders of magnitude larger than the 1 kHz homogeneous broadening predicted before. While the transition frequencies shift a couple of MHz with temperature \cite{higgins2025temperature}, the linewidths remain essentially unchanged over the investigated range ($\sim$150--293 K). In contrast, the linewidths depend strongly on both the quadrupole state and thorium concentration. Analysis based on the quadrupole-interaction Hamiltonian identifies static variations of the local electric-field gradient (EFG) as the dominant source of inhomogeneous broadening. The extracted EFG variations scale linearly with thorium concentration and can be consistently explained  by long-range lattice distortions induced by the Th dopants. Extrapolation of the concentration-dependent linewidths reveals a common nonzero contribution across the different transitions, indicating that the residual broadening also originates predominantly from EFG disorder, likely associated with crystal vacancies and interstitials, i.e., non-Th impurities (rather than e.g., dislocation or grain-boundary) defects. We further show theoretically that the $1/r^3$ random  spatial strain dependence due to the dopant and impurity defect induced EFG perturbations naturally accounts for (i) the observed Lorentzian line shapes, (ii) the linear concentration dependence of the linewidths, and (iii) the extrapolation of the residual line width. Complementary atomistic and \textit{ab initio} density-functional-theory (DFT) calculations reproduce these characteristic features, in agreement with the theoretical model and provide microscopic insight into the origin of the broadening and the local structure of the Th dopants.

%\KH{Maybe cut down a little the length of the intro?}

\begin{figure*}[tb]
\centering
\includegraphics[width=0.9\linewidth]{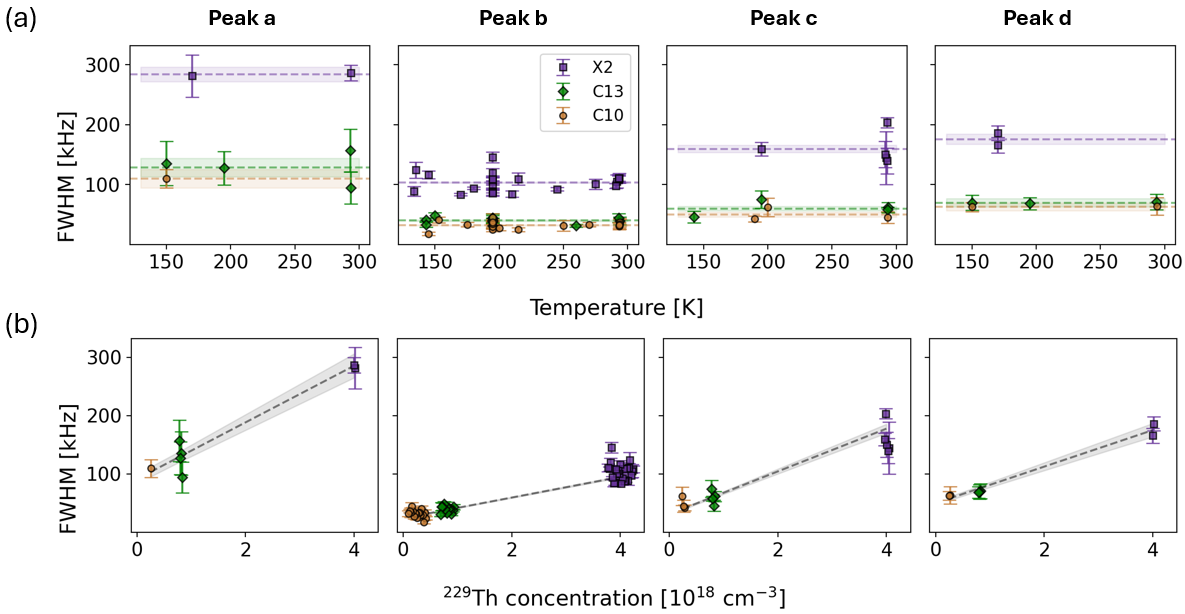}% Here is how to import EPS art
\caption{\label{fig:data} 
Linewidths of different $^{229}$Th nuclear transitions measured at various temperatures (a) and doping concentrations (b). Three crystals with different $^{229}$Th doping concentrations were investigated X2 ($\sim$$4\times 10^{18}$ \unit{cm^{-3}}), C13 ($\sim$$0.8 \times 10^{18}$ \unit{cm^{-3}}), C10 ($\sim$$0.3 \times 10^{18}$ \unit{cm^{-3}}). The linewidths differ among the transitions, and show no significant dependence on crystal temperature, but increase with increasing doping concentration.
}
\end{figure*}

%\section{\label{sec:level1} data and observations}
\section{Measurements}
To enable laser M\"ossbauer spectroscopy, $^{229}$Th nuclei are embedded in VUV-transparent crystalline hosts. In this work, 
$^{229}$Th-doped CaF$_2$ single crystals were grown at TU Wien using the vertical-gradient-freeze method \cite{beeks2023growth, beeks2024optical}. The large band gap of CaF$_2$ ($\sim$12 eV) exceeds the nuclear transition energy, largely preventing VUV absorption by the host crystal and suppressing the $^{229}$Th nonradiative decay channels \cite{beeks2023growth}. The thorium ions must be in the 3+ or higher charge state to suppress internal-conversion decay. Extensive theoretical and experimental studies indicate that thorium is incorporated predominantly as Th$^{4+}$ substituting for Ca$^{2+}$ lattice sites \cite{dessovic2014229}. Several microscopic defect configurations have been proposed. As illustrated in Fig.~\ref{fig:transitions}(a), two neighboring substitutional Th$^{4+}$ ions can form a dimer configuration, whose local charge distribution generates an EFG at the Th nuclei and produces the nuclear quadrupole splitting. Alternative configurations involve charge compensation by two fluorine interstitials, or by a Ca$^{2+}$ vacancy. The detailed microscopic structure of the doping site and its charge-compensation geometry remain subjects of active investigation \cite{pimon2022ab, takatori2025characterization,Rehn2026ThCaF2DefectAssignment}.

The transition frequency of $^{229}$Th in a crystal is influenced by its local electronic environment. The electron density at the nucleus produces an isomeric shift because the ground and isomeric states possess different nuclear charge radii. Simultaneously, the surrounding charge distribution may generate a local EFG, which couples to the nuclear quadrupole moment and lifts the degeneracy of the nuclear spin sublevels. The ground state and excited state nuclear spins are $I=5/2$ and $3/2$, respectively, and the five allowed transitions given by the selection rule $\Delta m =0$ or $\pm 1$ are shown in Fig.~\ref{fig:transitions}(b). The $^{229}$Th transition energy is described by the isomeric shift and quadrupole interaction Hamiltonians: %\KH{Make it explicit that R depends on GS vs IS. Otherwise it might be unclear how it couples to the transition energy. For (2) it is clear via $I_{x,y,z}$}:
\begin{equation}
H_0=\frac{2}{5}\pi Z e^2 R^2 \rho_e(0),
\label{eq:isomeric_shift}
\end{equation}
\begin{equation}
H_1=\frac{{Q_I} V_{zz}}{4I(2I-1)}
\left[3I_z^2-\boldsymbol{I}^2+\eta(I_x^2-I_y^2)\right],
\label{eq:qudrople_splitting}
\end{equation}
%{\color{red} LR: maybe it is an obvious point (for practitioners) but at the technical level above (1) and (2) are missing 8.4 eV GS-IS splitting. Interested by uninitiated reader may ask "where is that splitting discussed in the introduction?"}
where $Z$ is the nuclear charge number, $e$ is the electron charge, and $\rho_e(0)$ is the electron number density at the nucleus. The nuclear radius $R$ depends on the nuclear state, taking the values $R_g$ and $R_{\rm is}$ for the ground and isomeric states, respectively. The difference in $H_0$ between these two states therefore shifts the nuclear transition energy, giving rise to the isomer shift. In Eq.~(\ref{eq:qudrople_splitting}), $Q_I$ 
%{\color{red} LR @Kai: or perhaps we just absorb e into Q in both places, unless it conflicts with  the standard notation; I see in H0 our $\rho_e$ also has no e inside it. So then perhaps makes sense to display e explicitly in both H1 and H0...OR absorb e inside both hamiltonians.} \rev{KL: That makes sense. I absorbed the $e$ into the nuclear moments and changed the Appendix accordingly.} 
denotes the spectroscopic nuclear quadrupole moment  of the ground state ($I=5/2$) and isomeric ($I=3/2$) states, $I$ is the nuclear spin, $V_{zz}$ is the principal component of the EFG tensor, and $\eta \equiv (V_{xx}-V_{yy})/V_{zz}$ characterizes the EFG asymmetry. These Hamiltonians describe the shifts and splittings induced by the local electronic environment and EFG, superimposed on the bare ground-state isomer transition energy of approximately 8.4~eV. 

We measured the $^{229}$Th nuclear clock transitions using VUV frequency-comb spectroscopy. The excitation source has a linewidth of approximately 1 kHz, and the nuclear fluorescence following resonant excitation is detected (see Refs.~\cite{zhang2024frequency,higgins2025temperature,ooi2026frequency} for experimental details). Four of the five allowed optical transitions were measured. The fifth transition $m=\pm 1/2 \rightarrow m=\pm 3/2$ has relatively weaker transition strength due to small Clebsch-Gordan coefficients \cite{kazakov2012performance} and was not scanned in this study. Representative spectra acquired at 195\,K with a thorium-229 concentration of $0.8 \times 10^{18}$ \unit{cm^{-3}}  are shown in Fig.~\ref{fig:transitions}(c). The resonances are well fitted with a Lorentzian profile, and the full widths at half maximum (FWHM) exhibit a strong dependence on the quadrupole states.

We systematically characterized the frequencies and linewidths of the $^{229}$Th nuclear transitions as functions of temperature and thorium concentration using three crystals X2 ($\sim$$4 \times 10^{18}$ \unit{cm^{-3}}), C13 ($\sim$$0.8 \times 10^{18}$ \unit{cm^{-3}}), C10 ($\sim$$0.3 \times 10^{18}$ \unit{cm^{-3}}), illustrated in Fig.~\ref{fig:transitions} and~\ref{fig:data}. The measurements reveal two key observations: (i) the transition frequencies shift with temperature (see \cite{higgins2025temperature}) 
%{\color{red} LR: we make this point many times and indeed it is very important but as far as I can tell in this paper this information on transition frequencies (as opposed to widths) is not given, because it is in your earlier paper. We thus should say this explicitly and give the reference} 
while the linewidths remain unchanged, and (ii) the transition frequencies are largely independent of thorium concentration, whereas the linewidths increase linearly with concentration. These contrasting behaviors suggest distinct physical mechanisms governing the transition energies and linewidth broadening.

\begin{figure*}[tb]
\centering
\includegraphics[width=0.9\linewidth]{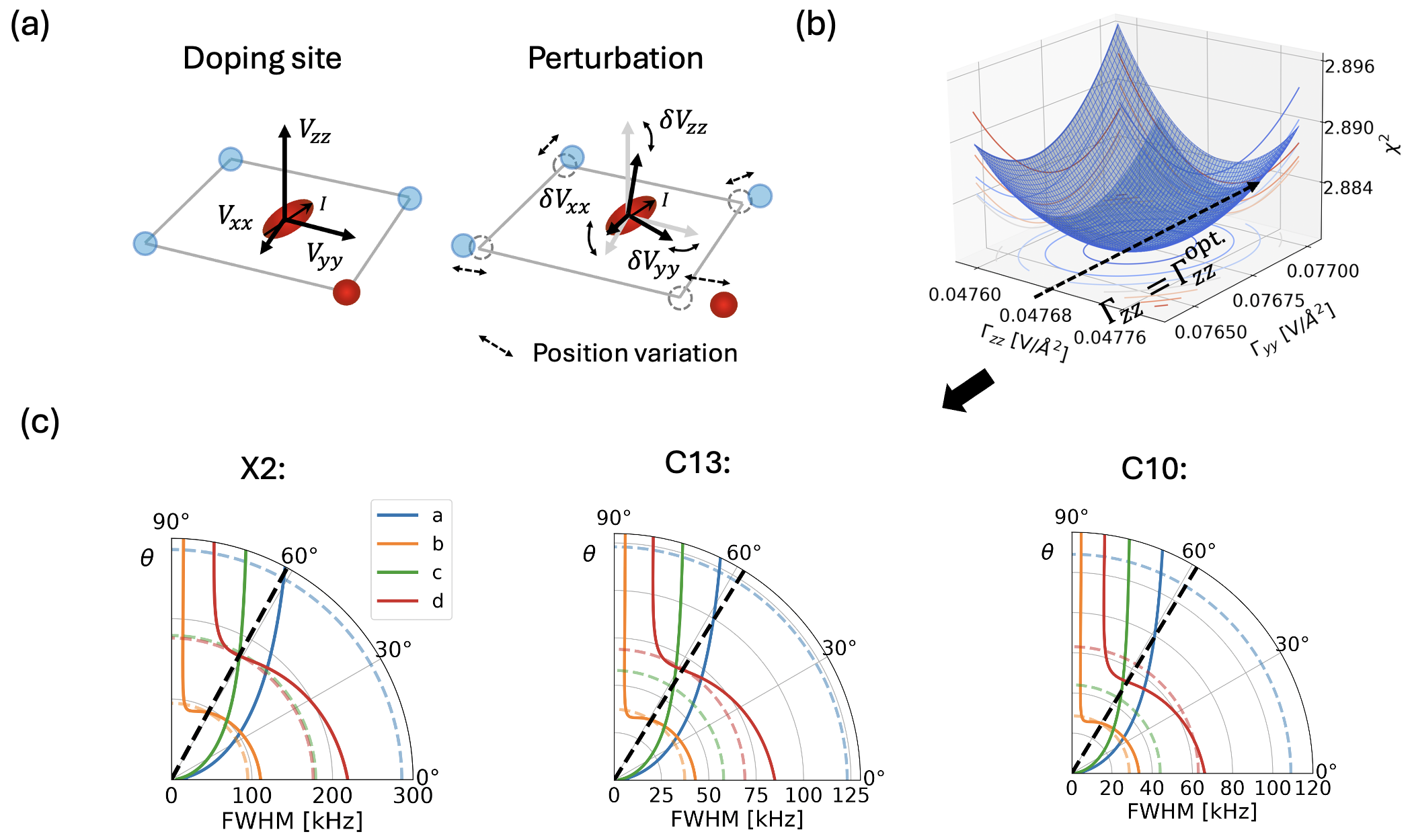}% Here is how to import EPS art
\caption{\label{fig:fit}
The schematic of inhomogeneous broadening mechanism and fitting results with coupling $\rho=-0.6$. (a) Schematic illustration of the local EFG at a $^{229}$Th doping site in CaF$_2$. Lattice distortions modify the positions of the surrounding ions, leading to variations of the EFG components and consequently to inhomogeneous broadening of the nuclear transitions. 
(b) Determination of the EFG variations $\Gamma_{zz}$ and $\Gamma_{yy}$. The linewidths of the individual transitions are calculated from the quadrupole-interaction Hamiltonian [Eq.~(\ref{eq:qudrople_splitting})], and the optimal EFG variations are extracted by minimizing the uncertainty-weighted $\chi^2$ between the calculated and measured linewidths.
(c) Comparison of measured linewidths and calculated values from Eq.~(\ref{eq:FWHM}) along the optimized $\Gamma_{zz}^\mathrm{opt.}$ (black dashed line in panel (b)). The measured linewidths are represented by the dashed lines at constant radius and the linewidths predicted by EFG variation model are represented by the solid lines. The dark dashed radial line indicates the best fit value of the angle $\theta$ defined by $\tan\theta=\Gamma_{yy}/\Gamma_{zz}$.
}
\end{figure*}

The temperature-dependent transition frequency shifts arise from two contributions: the isomeric (monopole) shift and the quadrupole splitting. Thermal expansion modifies the crystal lattice and thereby changes both the electronic charge density $\rho_e(0)$ at the nucleus [Eq.~(\ref{eq:isomeric_shift})] and the local EFG [Eq.~(\ref{eq:qudrople_splitting})].
The monopole contribution can be isolated by calculating the weighted average $f_{\mathrm{unsplit}}=\frac{1}{6}(f_a+2f_b+2f_c+f_d)$, which removes the quadrupole splitting.  Across the investigated temperature range ($\sim$150--293 K), $f_{\mathrm{unsplit}}$ changes by approximately $780$ kHz. The corresponding fractional volume change is $\Delta V/V=c\Delta T\simeq0.0025$, where $c$ is the thermal-expansion coefficient of CaF$_2$. For small density change caused by %[{\color{red} LR: a bit of an issue of confusion of $V$ volume and $V_{ij}$ EFG. Probably OK, but not optimum, esp when appearing in the same paragraph.}] 
$\Delta V$, $\Delta f_{\mathrm{unsplit}}/f_{\mathrm{iso}}\simeq\Delta V/V$, and we estimate an isomeric shift of $f_{\mathrm{iso}}\sim0.3$ GHz for $^{229}$Th embedded in CaF$_2$ relative to the bare-nucleus transition, consistent with previous theoretical predictions at the gigahertz level~\cite{kazakov2012performance}. The quadrupole contribution is then obtained by subtracting the common isomeric shift from each transition frequency. The EFG tensor responsible for the quadrupole interaction is traceless in the absence of free electric charges (i.e., with the monopole contribution removed). It therefore satisfies Laplace's equation, \(V_{xx}+V_{yy}+V_{zz}=0\), leaving only two independent parameters, which we take to be \(V_{zz}\) and \(V_{yy}\), equivalent to the two free parameters $V_{zz}$ and $\eta \equiv (V_{xx}-V_{yy})/V_{zz}$ shown in quadrupole splitting Hamiltonian Eq. (\ref{eq:qudrople_splitting}). From the measured temperature-dependent shifts, we extract the EFG changes $\Delta V_{ii}=V_{ii}(150\, {\rm K})-V_{ii}(293\, {\rm K})$ required to reproduce the observed quadrupole shifts, obtaining a relative change $\Delta V_{zz}/V_{zz}\sim0.01$ and a ratio $\Delta V_{yy}/\Delta V_{zz}\sim -1.6$.

Importantly, despite the substantial temperature-induced frequency shifts, no measurable temperature dependence of the linewidth is observed, as shown in Fig.~\ref{fig:data}(a). This indicates that the dominant broadening mechanism is associated with the static local crystal environment disorder rather than thermally driven dynamics. One might expect the thermally excited phonons to have a significant effect on the linewidths, since phonons modulate the atomic positions and hence the isomeric shift and quadrupole interaction. However, phonons are at very high frequencies and their effects are almost completely averaged out (motional narrowing) during the time of measurements.

In contrast to the temperature dependence, increasing the $^{229}$Th concentration broadens the nuclear transitions while leaving their center frequencies essentially unchanged. The linewidth dependence on concentration is shown in Fig.~\ref{fig:data}(b), and the measured FWHM values for all transitions and crystals are summarized in Table~\ref{tab:table1}. For each transition, the linewidth increases linearly with thorium concentration. The corresponding slopes, which quantify the sensitivity of the linewidth to concentration, follow the ordering $a>c\approx d>b$, identical to the ordering of the measured linewidths themselves. This correlation suggests a common physical origin for both the linewidth magnitude and its concentration dependence. Furthermore, the strong transition-to-transition variation in linewidth indicates that the broadening mechanism is closely linked to the quadrupole interaction. Taken together, these observations point to inhomogeneous broadening arising from dopant-induced static variations of the local EFG, which modify the quadrupole splitting and generate a distribution of transition frequencies.

\begin{table}[h]%The best place to locate the table environment is directly after its first reference in text
\caption{\label{tab:table1}%
Summary of the measured linewidth and fitting results for different transitions in different crystals.
}
\begin{ruledtabular}
\begin{tabular}{lcccc}
\textrm{Crystal}&
\textrm{$a$}&
\multicolumn{1}{c}{\textrm{$b$}}&
\multicolumn{1}{c}{\textrm{$c$}}&
\textrm{$d$}\\
\colrule
X2 (kHz)   & 286(12) & 94.8(0.9) & 179(6) & 176(9)  \\
C13 (kHz)   & 123(15) & 37.3(1.0) & 58(3) & 69(7) \\
C10 (kHz)   & 109(15) & 28.6(1.1) & 44(4) & 63(7)  \\
Intercept (kHz) & 91(13)  & 23.6(0.9) & 31(3)  & 50(6) \\
Slope (\unit{kHz/10^{18}cm^{-3}})  & 48(5)  & 17.8(0.3) & 37(2)  &31(3) \\

\end{tabular}
\end{ruledtabular}
\label{table:measured}
\end{table}

%\section{\label{sec:level1} Fitting}
\section{Data analysis}
The observed broadening is consistent with an inhomogeneous mechanism arising from variations in the local crystal environment. As illustrated in Fig.~\ref{fig:fit}(a), different nuclei experience slightly different EFGs, characterized by variations in $V_{zz}$ and $V_{yy}$, resulting in a distribution of nuclear transition frequencies. We assume that the underlying EFG variations $\delta V_{zz}$ and $\delta V_{yy}$ follow an elliptical bivariate Cauchy distribution, whose marginal distributions along $z$ and $y$ principle axis are Lorentzian as predicted by theory (see Sec. \ref{sec:theory}). For small EFG variations, the transition frequency can be linearized with respect to $\delta V_{zz}$ and $\delta V_{yy}$ and the resulting transition frequency distribution is a Lorentzian with width of (see Appendix~\ref{app:model_comparison})
\begin{equation}
\begin{split}
\Gamma_f=
\left[
\left(\frac{\partial f}{\partial V_{zz}}\right)^2\Gamma_{zz}^2+
\left(\frac{\partial f}{\partial V_{yy}}\right)^2\Gamma_{yy}^2 \right. \\ \left.
+2\rho
\frac{\partial f}{\partial V_{zz}}
\frac{\partial f}{\partial V_{yy}}
\Gamma_{zz}\Gamma_{yy}
\right]^{1/2},
\end{split}
\label{eq:FWHM}
\end{equation}
where $\Gamma_{ii}$ denotes the width of the marginal Lorentzian distribution of the EFG component $\delta V_{ii}$, and $\rho$ is a dimensionless parameter describing the coupling and orientation of the elliptical bivariate Cauchy distribution.

The transition frequency $f$ is calculated from the eigenvalues of the Hamiltonian $H_1$, Eq.~(\ref{eq:qudrople_splitting}) \cite{zory1965nuclear,gerdau1969quadrupole} (see also Appendix~\ref{app:H_EFG}). The derivatives are evaluated at the reference values $V_{zz}^0=109.1$ \unit{V/\angstrom^2} and $V_{yy}^0=-86.8$ \unit{V/\angstrom^2}, as determined from previous measurements \cite{zhang2024frequency}. The optimized parameters $\Gamma_{zz}$, $\Gamma_{yy}$, and $\rho$ for each crystal are obtained by minimizing the uncertainty-weighted $\chi^2$, as shown in Fig.~\ref{fig:fit}(b).

We first considered models with coupling $\rho=0$ and $\rho=1$. Both yield comparable poor fits, as quantified by the reduced chi-squared $\chi_{\rm red}^2$ (see Appendix~\ref{app:model_comparison}). Allowing $\rho$ to vary as an additional fitting parameter gives optimized values $\rho=[-0.60,-0.56,0.06]$ for the three crystals and substantially reduces $\chi_{\rm red}^2$ toward unity, demonstrating that coupling between the variations $\delta V_{zz}$ and $\delta V_{yy}$ is important for reproducing the measured transition-dependent linewidths. Because two of the three crystals independently yield $\rho\simeq-0.6$, we fix $\rho=-0.6$ in Eq.~(\ref{eq:FWHM}) and optimize $\Gamma_{zz}$ and $\Gamma_{yy}$ for each crystal.

The calculated linewidths are displayed in Fig.~\ref{fig:fit}(c) as polar plots for the optimized $\Gamma_{zz}^{\mathrm{opt.}}$, where the radius represents the linewidth and the angle is defined by $\tan\theta=\Gamma_{yy}/\Gamma_{zz}$. Overall, the model reproduces the measured linewidths remarkably well. Slight deviations are observed for transitions $a$ and $d$, likely reflecting the larger experimental uncertainties associated with the limited number of measurements available for these transitions. Notably, the optimized directions are nearly identical for the three crystals, with $\theta=[61^\circ,58^\circ,59^\circ]$, corresponding to $\Gamma_{yy}/\Gamma_{zz}=[1.8,1.6,1.6]$. The consistency of this ratio across crystals points to a characteristic value of $\Gamma_{yy}/\Gamma_{zz}\sim1.6$ with $\rho=-0.6$, whose magnitude is close to the ratio $\Delta V_{yy}/\Delta V_{zz}\sim-1.6$ inferred from the temperature-dependent EFG shifts discussed above. Although the two describe different aspects of the EFG response, the similar magnitudes suggest that the static EFG disorder and the temperature-induced EFG variation may reflect a common underlying microscopic response of the local dopant environment.

\begin{figure}[!htb]
\centering
\includegraphics[width=0.9\linewidth]{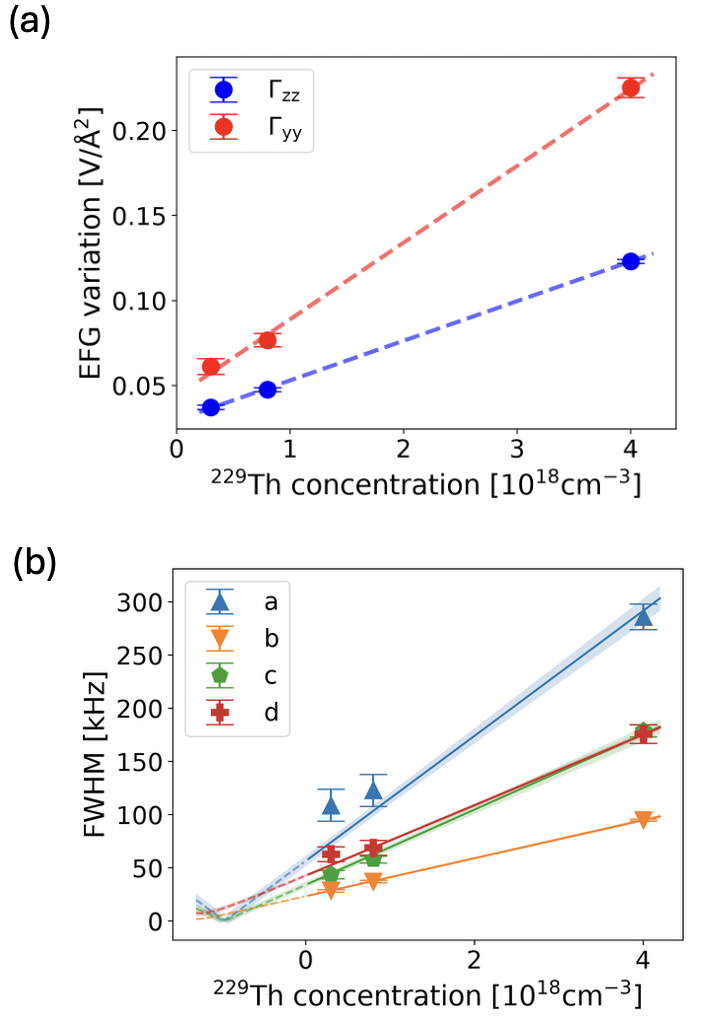}% Here is how to import EPS art
\caption{\label{fig:fit_redults}
The fitting results with coupling $\rho=-0.6$. (a) Optimized EFG variations $\Gamma_{zz}$ and $\Gamma_{yy}$ obtained from the fit. Both scale linearly with the $^{229}$Th concentration.
(b) Comparison of measured and model predicted linewidths. Extrapolation the linewidth vs.\ concentration shows that the linewidths of all four transitions converge to a common point, indicating the zero-doping broadening is also arising from EFG disorder in the host crystal.
}
\end{figure}

The optimized values of $\Gamma_{zz}$ and $\Gamma_{yy}$ extracted from the fit are plotted in Fig.~\ref{fig:fit_redults}(a) as functions of the $^{229}$Th concentration. The finite intercepts at zero concentration indicate an additional thorium concentration-independent contribution to the EFG disorder. The total EFG variation therefore comes from two sources: 
Th dopant induced lattice distortions $\delta\mathrm{EFG}_{\mathrm{Th-Th}}$ and the residual disorder $\delta\mathrm{EFG}_{\mathrm{crystal}}$ in the host crystal. Importantly, the measured lineshapes are well described by Lorentzian rather than Gaussian or Voigt profiles across the investigated concentration range, suggesting that the zero-concentration broadening contribution is also Lorentzian. As presented in Sec.~\ref{sec:theory} Theory, a Lorentzian distribution arises naturally from EFG perturbations with a long-range $1/r^3$ dependence. This suggests that the nonzero linewidth at vanishing thorium concentration comes from similar impurity sources of EFG perturbations with $1/r^3$ long-range behavior. This is consistent with attributing the zero-concentration EFG disorder to residual strain and structural imperfections of the host crystal.  

A direct comparison between the calculated and measured linewidths is presented in Fig.~\ref{fig:fit_redults}(b). The calculated FWHM values are obtained from Eq.~(\ref{eq:FWHM}) using the linearly fitted $\Gamma_{zz}$ and $\Gamma_{yy}$. Overall, the model reproduces the measured linewidths across all crystals and transitions. Notably, the linewidths remain finite in the zero-doping limit. Extrapolating the linewidths beyond the measured concentration range reveals that all four transitions converge to a common crossing point. This behavior indicates that the zero-doping linewidths originate from a common mechanism, namely the residual crystal EFG disorder, $\delta \mathrm{EFG}_{\mathrm{crystal}}$, associated with imperfections of the host crystal. 
%and thus the FWHM of line $i$ can be characterized by $\Gamma_i = \alpha_i(n_\text{Th} + n_\text{imp})$. 
Moreover, FWHM of the common crossing point remains finite, suggesting an additional broadening contribution at the $\sim$5 kHz level even after accounting for EFG-induced inhomogeneous broadening. Because this residual linewidth is transition-independent, it is most naturally attributed to an inhomogeneous broadening from the isomeric (monopole) shift Eq.~(\ref{eq:isomeric_shift}) and homogeneous broadening mechanism at the 1 kHz level expected from magnetic dipole interactions plus the second-order Doppler effect \cite{rellergert2010constraining,kazakov2012performance}. If both broadening mechanisms give rise to Lorentzian lineshapes, our assumption in extrapolating the linewidth to zero doping density—namely, that the linewidth contributions from the two mechanisms can be simply added—is justified \emph{a posteriori}.

%\section{\label{sec:level1} Discussion}

\begin{figure*}[tb]
\centering
\includegraphics[width=0.9\linewidth]{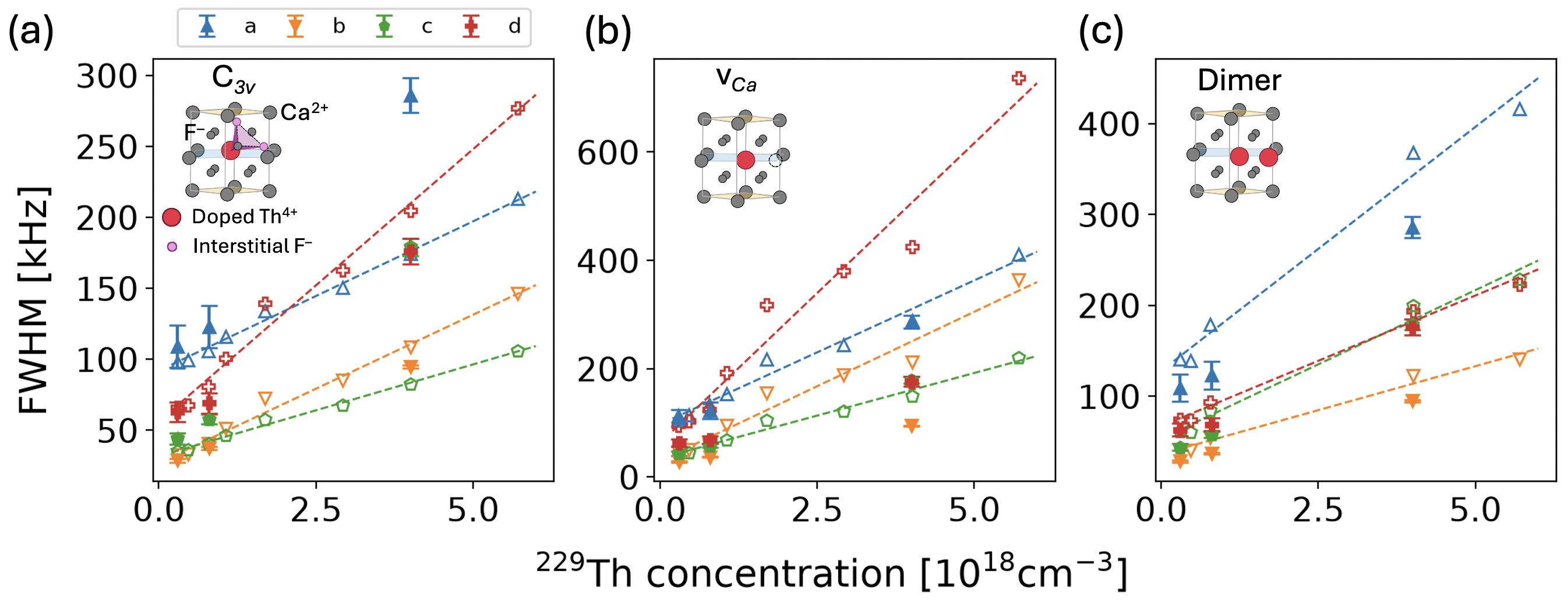}% Here is how to import EPS art
\caption{Linewidths calculated from \textit{ab initio} theory as a function of $^{229}$Th concentration for the (a) $\mathrm{C}_{3v}$, (b) $\mathrm{v}_{\mathrm{Ca}}$, and (c) dimer configurations. Experimental (filled) and theory (open) data points share the same symbol for each quadrupole component, the dotted lines are linear fits to theory. Among the three configurations, only the dimer reproduces the experimentally observed hierarchy of concentration-dependent transition linewidths, although the calculated linewidths are approximately an order of magnitude larger than the measured values and are therefore scaled by a factor of $\sim$0.06.
\label{fig:fit_theory}
}
\end{figure*}

\section{\label{sec:theory}Theory}
The experimentally observed linear dependence of linewidth on $^{229}$Th concentration can be understood from the long-range dopant-induced perturbations. Two mechanisms contribute to the local EFG disorder. First, a charged defect produces an EFG that decays asymptotically as $r^{-3}$. Second, a substitutional defect, vacancy, or interstitial acts at long distances as an elastic force dipole, generating a strain field that decays as $r^{-3}$ (Appendix~\ref{app:sources}). For a random thorium dopant density $n$, the characteristic separation between neighboring dopants scales as $d\sim n^{-1/3}$. Thus, the frequency shift has a characteristic scale $s\sim A/d^3\propto n$, which naturally explains the experimentally observed linear dependence of the linewidth on $n$. Here, $A$ characterizes the strength of the dopant-induced perturbation and depends on the microscopic defect geometry, local charge compensation, and screening (Appendix~\ref{app:sources}).

The same long-range interaction also accounts for the observed Lorentzian, rather than Gaussian, line shape. For randomly distributed dopants obeying Poisson statistics, the nearest-neighbor distance follows $P(r)dr=4\pi nr^2\,\exp(-4\pi nr^3/3)dr$. Transforming from distance to frequency shift gives $P(s)=s_0 s^{-2}\,\exp(-s_0/s)$, where $s_0 \equiv 4\pi A n/3$. In the large-shift limit, $P(s)\propto s^{-2}$, reproducing the characteristic heavy tail of a Lorentzian distribution (\cite{Stoneham_1969} and Appendix~\ref{app:nearest_tail}). Extending this nearest-neighbor argument to include the cumulative contributions from all randomly distributed dopants yields a stable distribution with characteristic function $\hat P(k)=\exp(-\Gamma|k|)$ and hence $\tilde P(s)=\Gamma/[\pi(s^2+\Gamma^2)]$, with $\Gamma\propto A n$, as derived in Appendix~\ref{app:detailed_lorentzian}. Thus, the $1/r^3$ scaling of the dopant-induced perturbation naturally accounts for two key experimental observations and corresponding simulations (discussed below and displayed in Fig.~\ref{fig:fit_theory}): the Lorentzian line shape and the linear dependence of the linewidth on thorium doping concentration.

To gain insights on the microscopic origin of the linewidth and the dopant structure, we combine classical atomistic simulations with quantum-mechanical \textit{ab initio} DFT calculations. Direct first-principles simulations at the experimental doping densities would require prohibitively large supercells. We therefore use classical atomistic calculations to determine the distribution of dopant-induced atomic displacements and microstrains, while DFT calculations provide the local response of the EFG to these displacements. The resulting EFG distributions are propagated through the nuclear quadrupole Hamiltonian to obtain distributions of transition frequencies and the corresponding spectra (Appendix~\ref{app:ab_initio_theory}).

As shown in Fig.~\ref{fig:fit_theory}, we consider three candidate defect configurations: $\mathrm{C}_{3v}$, $\mathrm{v}_{\mathrm{Ca}}$, and dimer. In the $\mathrm{C}_{3v}$ configuration, Th substitutes for Ca and is charge compensated by two interstitial fluorine ions. Structural relaxation from the unstable $\mathrm{C}_{2v}$-symmetric configuration with a $90^\circ$ F--Th--F geometry yields a $\mathrm{C}_{3v}$-symmetric configuration. The $\mathrm{v}_{\mathrm{Ca}}$ configuration consists of substitutional Th accompanied by a neighboring Ca vacancy. Finally, the dimer configuration consists of two Th ions occupying nearest-neighbor Ca sites with nonlocal charge compensation \cite{Dessovic2014ThCaF2,Takatori2025ThCaF2DefectStructure,Pimon2022CaF2Defects,Hiraki2026LaserMossbauer,Rehn2026ThCaF2DefectAssignment}.

For all three configurations, the simulated frequency distributions exhibit Lorentzian line shapes and linewidths that increase linearly with dopant concentration, confirming that these features follow generically from the long-range dopant-induced perturbations. The DFT-simulated equilibrium EFGs, however, provide a discriminator among the microscopic configurations. In a $2 \times 2 \times 2$ supercell of the conventional CaF$_2$ unit cell, the $\mathrm{C}_{3v}$ and $\mathrm{v}_{\mathrm{Ca}}$ structures yield $(V_{zz},\eta)=(-79.4~\mathrm{V/\AA^2},0)$ and $(-101.57~\mathrm{V/\AA^2},0.165)$, respectively, both substantially different from the experimentally determined values $V_{zz}=109.1(7)~\mathrm{V/\AA^2}$ and $\eta=0.59163(5)$. In contrast, the dimer configuration gives $V_{zz}=92.56~\mathrm{V/\AA^2}$ and $\eta=0.861$, closer to experiment. Thus, among the configurations considered here, the dimer provides the most consistent description of the measured local EFG.

To isolate the broadening physics from discrepancies in the calculated equilibrium EFGs, we place all three configurations on equal footing by fixing their mean EFG to the experimentally measured values, $V_{zz}=109.1(7)~\mathrm{V/\AA^2}$ and $\eta=0.59163(5)$ and added the experimental intercepts, while retaining the simulated EFG fluctuations. The resulting linewidths are compared with experiment in Fig.~\ref{fig:fit_theory}. Both the $\mathrm{C}_{3v}$ and $\mathrm{v}_{\mathrm{Ca}}$ configurations predict a hierarchy of transition linewidths inconsistent with experiment, whereas the dimer reproduces the observed ordering. Applying the same EFG-fluctuation analysis used for the experimental data yields $\Gamma_{yy}/\Gamma_{zz}\sim0.6$ for both the $\mathrm{C}_{3v}$ and $\mathrm{v}_{\mathrm{Ca}}$ configurations. For the dimer, assuming uncoupled variations (using Eq.~(\ref{eq:FWHM}) with $\rho=0$), the fitting gives $\Gamma_{yy}/\Gamma_{zz}\sim3.1$. 
Fitting with coupled linewidth model Eq.~(\ref{eq:FWHM}) yields $\Gamma_{yy}/\Gamma_{zz}\sim1.6$ and coupling $\rho\approx -1$ (see Appendix \ref{app:ab_initio_theory}), in substantially better agreement with the experimentally observed ratio than other configurations and consistent with the thermal expansion determined $\Delta V_{yy}/\Delta V_{zz}\approx -1.6$. This sensitivity to correlations indicates that correlated EFG fluctuations play an important role in the dimer configuration. Such correlations can be physically understood as arising from the uncompensated charge of the Th dimer, which generates strong Coulomb interactions with the surrounding lattice, leading to coupled atomic displacements and, consequently, correlated fluctuations of the EFG components.

The dimer configuration therefore uniquely reproduces several independent experimental signatures: (i) an equilibrium EFG closest to the measured value, (ii) the observed hierarchy of transition linewidths, and (iii) an EFG-fluctuation ratio comparable to experiment. Quantitatively, however, the calculated dimer linewidths are approximately an order of magnitude larger than measured (Fig.~\ref{fig:fit_theory} (c) is scaled by a factor of 0.06), whereas the $\mathrm{C}_{3v}$ and $\mathrm{v}_{\mathrm{Ca}}$ configurations produce broadening at the correct scale. In the simulations, the excessive dimer broadening originates from correspondingly larger atomic displacements driven by the long-range Coulomb interaction of the charged dopants. This discrepancy suggests that the actual crystal may contain additional screening or nonlocal charge compensation not captured by the present model. Further microscopic modeling of charge compensation and screening will therefore be necessary for a quantitative description of the linewidth.

\section{Discussion}
Solid-state $^{229}$Th nuclear clocks benefit from the substantial signal enhancement provided by embedding a macroscopic number of nuclei in a crystalline host. However, their performance is presently limited by transition linewidths that far exceed the natural linewidth set by the radiative lifetime of the nuclear isomer. Understanding the origin of this broadening is therefore essential for advancing solid-state nuclear clocks. In this work, we systematically characterized the linewidths of different quadrupole transitions as functions of temperature and thorium concentration. We find that the linewidth is independent of temperature, increases linearly with thorium concentration, and depends strongly on the quadrupole state. These observations point to an inhomogeneous broadening mechanism arising from static variations of the local EFG, which modify the quadrupole splitting. By fitting the measured linewidths with a quadrupole-interaction Hamiltonian, we show that the extracted EFG variations scale linearly with thorium concentration. Furthermore, extrapolation of the linewidths vs.\ concentration reveals a common intercept for all transitions, indicating that the residual linewidth is also dominated by EFG disorder. We note that the monopole and quadrupole shifts may also be correlated through their common dependence on lattice distortions, potentially modifying their combined contribution to the linewidth and the fitting model. However, such correlations cannot be independently constrained by the present data, and their quantitative investigation is left to future theoretical and experimental studies. Although the present analysis employs a simplified EFG model, it successfully reproduces the observed Lorentzian line shape and measured linewidths across all crystals and transitions and establishes EFG disorder as the dominant broadening mechanism. 

A notable result is the nearly constant ratio $\Gamma_{yy}/\Gamma_{zz} \sim 1.6$ with correlation $\rho=-0.6$ obtained for all three crystals, suggesting a common microscopic origin of the EFG variations ratio. The detailed charge-compensation geometry of the $^{229}$Th doping site remains an active area of investigation, and the observed ratio provides an important experimental constraint. The \textit{ab initio} DFT calculations of Th-dimer configuration gives the ratio $\sim$ 1.6 with $\rho \sim -1$, comparable to the experimentally extracted value and the change $\Delta V_{yy}/\Delta V_{zz}=-1.6$ during thermal expansion, while the $\mathrm{C}_{3v}$ and $\mathrm{v}_{\mathrm{Ca}}$ configurations find ratio $\sim$ 0.6. The measured EFG variations, $\Gamma_{zz}\sim0.05$ \unit{V/\angstrom^2}, correspond to atomic displacements on the order of $10^{-4}$\unit{\angstrom}. At a typical thorium concentration of $10^{18}$ \unit{cm^{-3}}, the average dopant separation is approximately $100$ \unit{\angstrom}, and classical atomistic simulations based on DFT calculations indicate that lattice distortions of this magnitude can be induced by neighboring thorium dopants.

The identification of EFG disorder as the dominant inhomogeneous broadening mechanism points to clear pathways for improving solid-state nuclear clock materials. Since the EFG variations appear to originate primarily from thorium-induced lattice distortions and residual crystal strain, reducing these sources of disorder should substantially narrow the nuclear resonances. One promising direction is the development of crystal hosts based on thorium compounds, such as ThF$_4$, which may eliminate dopant-induced lattice distortions altogether \cite{zhang2024229thf4}. Further improvements in crystal growth, purification, and annealing could reduce residual strains and suppress the zero-doping linewidth contribution. Within our linewidth model, the extrapolation places an upper bound of $\sim$5~kHz on the residual linewidth after EFG-induced inhomogeneous broadening is removed. This bound is consistent with the possibility of approaching the homogeneous-broadening regime, for which theoretical estimates of individual mechanisms, including magnetic dipole interactions and the second-order Doppler effect, are at the $\sim$1~kHz scale. Even before such low-disorder samples become available, homogeneous broadening may be probed using techniques that are insensitive to inhomogeneous disorder, including measurements of the nuclear forward-scattering coherence lifetime under narrowband VUV excitation or radio-frequency electromagnetically induced transparency spectroscopy. These approaches provide promising routes toward resolving the homogeneous linewidth of the $^{229}$Th nuclear transition and realizing the full potential of solid-state nuclear clocks.

\begin{acknowledgments}
Work at JILA is funded by the DOE Quantum Center of Quantum System Accelerator, DARPA (HR0011-25-3-0140), the Army Research Office (W911NF-25-1-0244), and the National Institute of Standards and Technology. LR and EP were supported by the Simons Investigator award (to LR) from the Simons Foundation. EP was also supported by the National Science Foundation Graduate Research Fellowship Program under Grant No. (DGE 2040434). This work utilized the Alpine High-Performance Computing resource at the University of Colorado Boulder. Alpine is jointly funded by the University of Colorado Boulder, the University of Colorado Anschutz, Colorado State University, and the National Science Foundation (award 2201538). SMG was supported by the U.S. Department of Energy, Office of Science, 
National Quantum Information Science Research Centers, Co-design Center for Quantum Advantage (C2QA) under contract number DE-SC0012704.
C2QA contributed to the theoretical analysis in this project. KH is supported by the Swiss National Science Foundation (SNSF) through Project Grant No. 239143.
Work in Vienna has been funded by the European Research Council (ERC) under the European Union’s Horizon 2020 research and innovation programme (Grant Agreement No. 856415,  No. 101266822 and No.  101087184) and the Austrian Science Fund (FWF) [Grant DOI: 10.55776/F1004, 10.55776/KIN4315625]. Parts of the computational results have been achieved using the Austrian Scientific Computing (ASC) infrastructure.
%\SMG{All to add any missing grant acknowledgments}
\end{acknowledgments}

\appendix

%%%%%%%%%%%%%%
\section{Comparison of fitting models}
\label{app:model_comparison}

\setcounter{table}{0}
\renewcommand{\thetable}{B\arabic{table}}

\begin{table*}[t]
\centering
\caption{Comparison of fits to the measured linewidths using different
models for the EFG variations. Here
$R\equiv\Gamma_{yy}/\Gamma_{zz}$, $\rho$ is the coupling parameter
of the elliptical bivariate Cauchy distribution, and
$\chi^2_{\rm red}$ is the reduced chi-squared.}
\label{tab:model_comparison}

\renewcommand{\arraystretch}{1.5}
\setlength{\tabcolsep}{5pt}

\begin{tabular}{l@{\hspace{20pt}}ccc|ccc|ccc}
\hline\hline
&
\multicolumn{3}{c|}{$R$}
&
\multicolumn{3}{c|}{$\rho$}
&
\multicolumn{3}{c}{$\chi^2_{\rm red}$}
\\
\cline{2-10}
Model
& X2 & C13 & C10
& X2 & C13 & C10
& X2 & C13 & C10
\\
\hline

$\rho=0$, Eq.~(\ref{eq:FWHM_rho_zero})
& 2.14 & 1.88 & 1.96
& 0 & 0 & 0
& 8.65 & 2.79 & 2.28
\\

$\rho=1$, Eq.~(\ref{eq:FWHM_rho_one})
& 2.90 & 2.35 & 2.47
& 1 & 1 & 1
& 26.54 & 6.35 & 2.83
\\

Fitted $\rho$, Eq.~(\ref{eq:cauchy_linewidth})
& 1.80 & 1.60 & 1.96
& $-0.60$ & $-0.56$ & 0.06
& 0.55 & 2.86 & 4.55
\\

Fixed $\rho=-0.6$, Eq.~(\ref{eq:cauchy_linewidth})
& 1.80 & 1.60 & 1.60
& $-0.60$ & $-0.60$ & $-0.60$
& 0.55 & 2.90 & 6.90
\\

\hline\hline
\end{tabular}
\end{table*}

The experimentally observed spectral lines are well described by
Lorentzian profiles, and the EFG distributions follow
Lorentzian marginals in theory. Unlike Gaussian distributions, Lorentzian
distributions have undefined second moments. Therefore, the EFG
variations cannot be characterized in terms of variances and
covariances. Instead, we characterize the distributions directly by
the FWHM, denoted by $\Gamma$.

We model the joint EFG fluctuations using an elliptical bivariate
Cauchy distribution,
with the normalized EFG variations
\begin{equation}
u\equiv\frac{\delta V_{zz}}{\Gamma_{zz}},
\qquad
v\equiv\frac{\delta V_{yy}}{\Gamma_{yy}}
\end{equation}

\begin{equation}
\begin{split}
p(\delta V_{zz},\delta V_{yy})
={}&
\frac{2}
{\pi\Gamma_{zz}\Gamma_{yy}\sqrt{1-\rho^2}}
\\
&\times
\left[
1+
\frac{4\left(u^2+v^2-2\rho uv\right)}
{1-\rho^2}
\right]^{-3/2}
\label{eq:bivariate_cauchy}
\end{split}
\end{equation}
where $\Gamma_{zz}$ and $\Gamma_{yy}$ are the FWHMs of the marginal
distributions of $\delta V_{zz}$ and $\delta V_{yy}$, respectively.
The dimensionless parameter $\rho$ determines the off-diagonal coupling
and orientation of the elliptical distribution. Because the moments of
a Cauchy distribution are undefined, $\rho$ should not be interpreted
as a conventional Pearson correlation coefficient.

Integrating Eq.~\eqref{eq:bivariate_cauchy} over either variable yields
the corresponding Lorentzian marginal distribution. For example,
\begin{equation}
\begin{split}
p_M(\delta V_{yy})
&=
\int_{-\infty}^{\infty}
p(\delta V_{zz},\delta V_{yy})\,d(\delta V_{zz})
\\
&=
\frac{\Gamma_{yy}}
{2\pi\left[
\delta V_{yy}^2+(\Gamma_{yy}/2)^2
\right]}
\label{eq:cauchy_marginal}
\end{split}
\end{equation}
which has width of distribution $\Gamma_{yy}$, with an analogous expression can be obtained for
$\delta V_{zz}$.

To first order in the EFG variations, the shift of a nuclear transition
frequency is
\begin{equation}
\delta f =
\frac{\partial f}{\partial V_{zz}}\delta V_{zz}
+
\frac{\partial f}{\partial V_{yy}}\delta V_{yy}
\label{eq:linear_EFG_shift}
\end{equation}
A linear projection of the elliptical bivariate Cauchy distribution
remains Cauchy. The FWHM of the resulting transition-frequency
distribution is therefore
\begin{equation}
\begin{split}
\Gamma_f =
\Bigg[
&
\left(
\frac{\partial f}{\partial V_{zz}}\Gamma_{zz}
\right)^2
+
\left(
\frac{\partial f}{\partial V_{yy}}\Gamma_{yy}
\right)^2
\\
&+
2\rho
\frac{\partial f}{\partial V_{zz}}
\frac{\partial f}{\partial V_{yy}}
\Gamma_{zz}\Gamma_{yy}
\Bigg]^{1/2}.
\label{eq:cauchy_linewidth}
\end{split}
\end{equation}

Equation~\eqref{eq:cauchy_linewidth} directly relates the FWHMs of the
underlying EFG distributions to the measured Lorentzian linewidths.
Importantly, $\rho=0$ does not imply statistical independence of
$\delta V_{zz}$ and $\delta V_{yy}$ within the elliptical bivariate
Cauchy model. In this limit,
\begin{equation}
\Gamma_f =
\left[
\left(
\frac{\partial f}{\partial V_{zz}}\Gamma_{zz}
\right)^2
+
\left(
\frac{\partial f}{\partial V_{yy}}\Gamma_{yy}
\right)^2
\right]^{1/2}
\label{eq:FWHM_rho_zero}
\end{equation}
whereas the widths of two statistically independent Cauchy variables
would add linearly under a linear combination. Thus, $\rho$
characterizes the off-diagonal coupling of the elliptical Cauchy
distribution rather than statistical independence.

To examine the role of this coupling in the measured linewidths, we fit
the data using several limiting and general cases of
Eq.~\eqref{eq:cauchy_linewidth}. We define
\begin{equation}
R \equiv \frac{\Gamma_{yy}}{\Gamma_{zz}},
\label{eq:R_definition}
\end{equation}
which characterizes the relative widths of the EFG variations in the
two principal components.

For $\rho=0$, Eq.~\eqref{eq:cauchy_linewidth} reduces to
Eq.~\eqref{eq:FWHM_rho_zero}. At the opposite limit, $\rho=1$, together
with $\Gamma_{yy}=R\Gamma_{zz}$, gives
\begin{equation}
\Gamma_f =
\left|
\left(
\frac{\partial f}{\partial V_{zz}}
+
R\frac{\partial f}{\partial V_{yy}}
\right)
\Gamma_{zz}
\right|.
\label{eq:FWHM_rho_one}
\end{equation}
We further consider the general case in which $\rho$ is either treated
as a free fitting parameter or fixed to a common value for all three
crystals.

The fitting results are summarized in
Table~\ref{tab:model_comparison}. The limiting cases $\rho=0$ and
$\rho=1$ provide comparatively poor descriptions of the measured
linewidths. Allowing $\rho$ to vary substantially improves the fit for
X2, reducing $\chi^2_{\rm red}$ from 8.65 for $\rho=0$ to 0.55. The
optimized coupling parameters are $\rho=-0.60$, $-0.56$, and $0.06$
for X2, C13, and C10, respectively. Thus, X2 and C13 independently
favor similar negative coupling parameters, whereas C10 shows no clear
preference for the same value.

Motivated by the similar values obtained independently for X2 and C13,
we also consider a common coupling parameter of $\rho=-0.6$ for all
three crystals. With $\rho$ fixed, the optimized width ratios are
$R=1.8$, $1.6$, and $1.6$ for X2, C13, and C10, respectively.
Although fixing $\rho=-0.6$ increases $\chi^2_{\rm red}$ for C10, the
resulting values of $\Gamma_{yy}/\Gamma_{zz}$ remain similar across the
three crystals. This consistency suggests that the relative widths of
the two EFG components may arise from a common microscopic mechanism,
while the overall magnitude of the EFG disorder varies among the
crystals.

\section{Spectrum of EFG Hamiltonian}
\label{app:H_EFG}

\renewcommand{\thefigure}{A\arabic{figure}}
\setcounter{figure}{0}

\begin{figure*}[tb]
\centering
\includegraphics[width=0.9\linewidth]{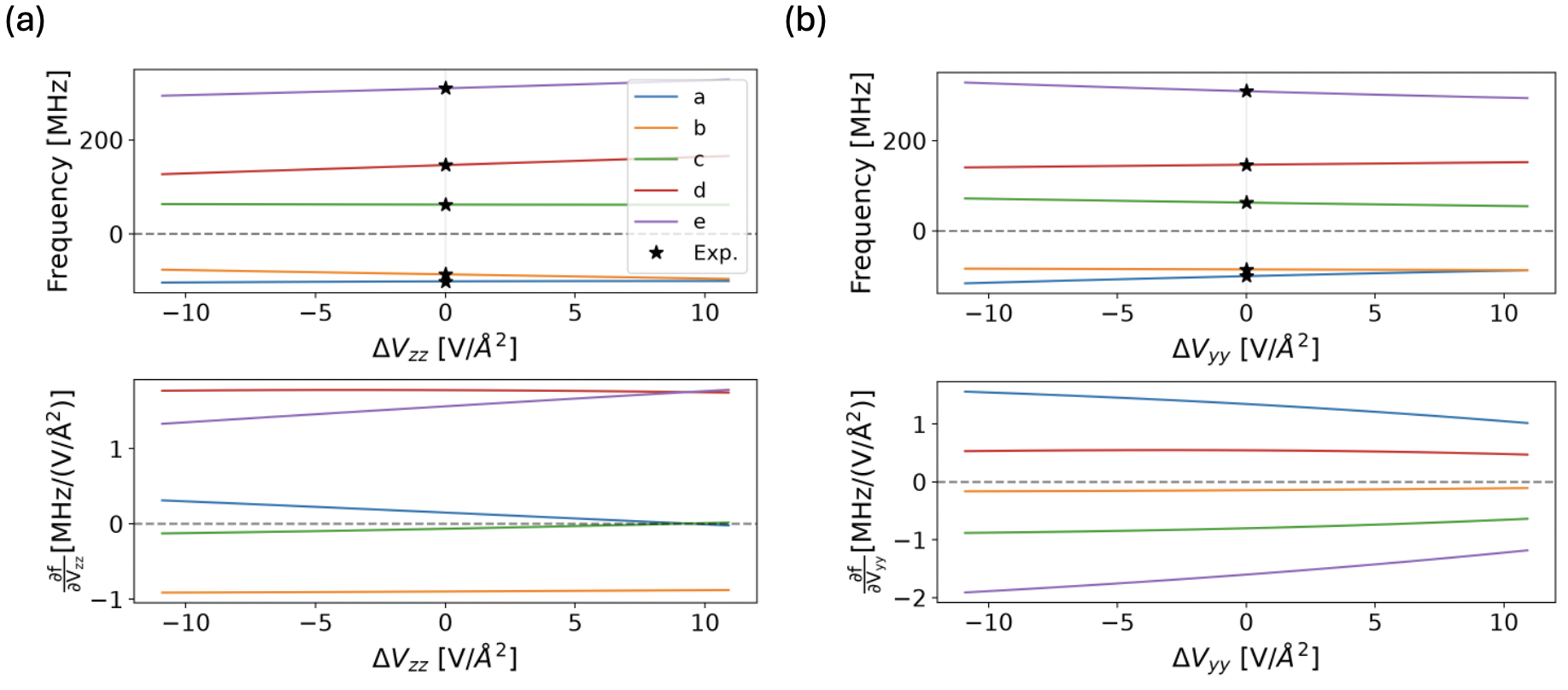}% Here is how to import EPS art
\caption{Transition frequencies and their sensitivities to variations in the EFG. The horizontal axes represent changes in the EFG components, $\Delta V_{zz}$ and $\Delta V_{yy}$, relative to the experimentally determined values $V_{zz}=109.1~\mathrm{V/\AA^2}$ and $V_{yy}=-86.82~\mathrm{V/\AA^2}$. (a) Upper panel: transition frequencies as a function of $\Delta V_{zz}$. Lower panel: the sensitivity of each transition frequency to $V_{zz}$, quantified by the partial derivative $\partial f/\partial V_{zz}$, as a function of $\Delta V_{zz}$ at fixed $V_{yy}=-86.82~\mathrm{V/\AA^2}$. (b) Same as (a), but showing the transition frequencies and their sensitivities to variations in $V_{yy}$ as functions of $\Delta V_{yy}$.
\label{fig:EFG_spec}
}
\end{figure*}

Here we give details of diagonalization of the quadrupole Hamiltonian $H_1$ controlling $^{229}$Th low-energy spectrum studied in the main text, illustrated in Fig.~\ref{fig:transitions}(b) and  Fig.~\ref{fig:EFG_spec}.
In the principal-axis frame of the EFG tensor, the nuclear quadrupole Hamiltonian is given by 
\begin{equation}
H_1 =
A_I\left[
3 I_z^2
%-I(I+1)
- \boldsymbol{I}^2
+\frac{\eta}{2}\left(I_+^2+I_-^2\right)
\right],
%\qquad
%A_I=\frac{eQ_I V_{zz}}{4I(2I-1)},
\label{eq:H1}
\end{equation}
where
\begin{equation}
A_I=\frac{Q_I V_{zz}}{4I(2I-1)},\;\; \eta=\frac{V_{xx}-V_{yy}}{V_{zz}},
\;\;
V_{xx}+V_{yy}+V_{zz}=0.
\label{eq:eta}
\end{equation}
The Hamiltonian is time-reversal invariant, and, for half-integer $I$,
the states form Kramers doublets.  The $\eta$ (in-plane biaxial) term couples
states with $\Delta m=\pm2$.  For $I=5/2$ the two time-reversed sectors
can be chosen as
\begin{equation}
\left\{
\left|\frac52\right\rangle,
\left|\frac12\right\rangle,
\left|-\frac32\right\rangle
\right\},
\qquad
{\cal T}
\left\{
\left|\frac52\right\rangle,
\left|\frac12\right\rangle,
\left|-\frac32\right\rangle
\right\}.
\label{eq:Kramerssectors}
\end{equation}

For the $^{229}$Th ground state, $I_g=5/2$, with
\begin{equation}
A_g=\frac{Q_gV_{zz}}{40},
\label{eq:Ag}
\end{equation}
the Hamiltonian in above basis is given by
\begin{equation}
H_g
=
A_g
\begin{pmatrix}
10 & \sqrt{10}\eta & 0\\
\sqrt{10}\eta & -8 & 3\sqrt{2}\eta\\
0 & 3\sqrt{2}\eta & -2
\end{pmatrix}
\otimes {\mathbf 1}_{K},
\label{eq:Hgmatrix}
\end{equation}
where ${\mathbf 1}_{K}$ is the $2\times 2$ identity matrix representing the Kramers degeneracy.
$H_g$ gives three energy eigenvalues, 
$E_{g,k}=A_g\lambda_k$, each Kramers degenerate, with $\lambda_k$ as the three roots of
\begin{equation}
\lambda^3
-28(3+\eta^2)\lambda
-160(1-\eta^2)=0,
\label{eq:groundcubic}
\end{equation}
or explicitly
\begin{widetext}
\begin{equation}
\lambda_k=
2\sqrt{\frac{28(3+\eta^2)}{3}}\,
\cos\left[
\frac{1}{3}
\cos^{-1}\left(
\frac{240\sqrt{3}(1-\eta^2)}
{\left[28(3+\eta^2)\right]^{3/2}}
\right)\nonumber
-\frac{2\pi k}{3}
\right],
\qquad k=0,1,2.
\label{eq:groundeigenvalues}
\end{equation}
\end{widetext}

For $\eta=0$, this reduces to 
\begin{equation}
E_g=A_g\{10,-2,-8\},
\label{eq:groundaxial}
\end{equation}
each with Kramers degeneracy, consistent with $H_1$ in (\ref{eq:H1}) automatically diagonal. 

For the isomer, $I_e=3/2$, with
\begin{equation}
A_e=\frac{Q_eV_{zz}}{12},
\label{eq:Ae}
\end{equation}
the two time-reversed sectors are
\begin{equation}
\left\{
\left|\frac32\right\rangle,
\left|-\frac12\right\rangle
\right\},
\qquad
\left\{
\left|-\frac32\right\rangle,
\left|\frac12\right\rangle
\right\}.
\label{eq:Kramerssectors32}
\end{equation}
giving an effectively pseudo-spin 1/2 form,
\begin{equation}
H_e
=
A_e
\begin{pmatrix}
3 & \sqrt{3}\eta\\
\sqrt{3}\eta & -3
\end{pmatrix}
\otimes {\mathbf 1}_{K},
\label{eq:Hematrix}
\end{equation}
and the two Kramers-degenerate energies are
\begin{equation}
E_{e,\pm}
=
\pm A_e\sqrt{9+3\eta^2}
=
\pm3A_e\sqrt{1+\frac{\eta^2}{3}}.
\label{eq:isomereigenvalues}
\end{equation}
Thus, the EFG Hamiltonian spectrum is given by
\begin{widetext} 
\begin{equation}
I_g=\frac52:\qquad
3\times3\otimes\text{Kramers doublet},
\qquad
I_e=\frac32:\qquad
2\times2\otimes\text{Kramers doublet},
\label{eq:spectrumsummary}
\end{equation}
as illustrated in Fig.~\ref{fig:transitions}(b) of the main text.
\end{widetext}

%\SMG{The caption of Fig.~\ref{fig:EFG_spec} does not specify the units for the vertical axis in the two lower panels. Maybe refer in the caption to the text below? Or move the numerical values below to the caption?  Maybe the best solution is to simply label the vertical axes in the lower panels with MHz/(V/$\AA^2$) if that is the correct unit.} \rev{KL: added the unit to the vertical axis. Thanks}
The transition frequencies, calculated from the energy differences between the corresponding nuclear states, are shown in Fig.~\ref{fig:EFG_spec}. Taking the partial derivatives of the transition frequencies with respect to the EFG components yields their sensitivities to variations in $V_{zz}$ and $V_{yy}$. These sensitivity coefficients are then used to fit the measured FWHMs according to Eq.~(\ref{eq:FWHM}) in the main text. At the experimental value $V_{yy}=-86.82~\mathrm{V/\AA^2}$, the sensitivities $\partial f/\partial V_{zz}$ for transitions $a$ to $e$ are $(0.1483,-0.8951,-0.0661,1.7742,1.5598)~\mathrm{MHz/(V/\AA^2)}$, respectively. Similarly, at the experimental value $V_{zz}=109.1~\mathrm{V/\AA^2}$, the corresponding sensitivities $\partial f/\partial V_{yy}=(1.3424,-0.1442,-0.7990,0.5439,-1.5975)~\mathrm{MHz/(V/\AA^2)}$.

\section{Lorentzian line shape with linear in-doping width}
\label{app:lorentzian}
\subsection{Sources of inhomogeneous broadening}
\label{app:sources}
Like all realistic solid state materials, CaF$_2$ crystals contain
vacancies, interstitials, dislocations, grain boundaries, and in the case of our samples,
substitutional thorium dopants.  These produce spatially varying local
EFGs and strains.  We focus on two mechanisms generated directly by Th
dopants and their compensating defects, because both have the same
long-distance $1/r^3$ random EFG perturbations and therefore the same
line-shape statistics.

\subsubsection{Random EFGs}
A charge $Ze$ at ${\bf R}$ produces the Coulomb potential at $\bf r$,
\begin{equation}
 V({\bf r})=\frac{Ze}{4\pi\epsilon|{\bf r}-{\bf R}|},
 \label{eq:coulomb_potential}
\end{equation}
where $\epsilon\approx 6.8$ is the dielectric constant of the CaF$_2$ host crystal.\cite{Andeen1971CaF2}
Its EFG at the observed Th nucleus, taken to be at ${\bf r}=0$, is
\begin{equation}
 \delta V_{\alpha\beta}({\bf R})
 =\left.\partial_\alpha\partial_\beta V({\bf r})\right|_{{\bf r}=0}
 =\frac{Ze}{4\pi\epsilon R^3}
 \left(3\hat R_\alpha\hat R_\beta-\delta_{\alpha\beta}\right).
 \label{eq:efg_point_charge}
\end{equation}
Thus every component of the EFG has the far-field scaling $\delta V_{\alpha\beta}\sim R^{-3}$.  Linearizing the quadrupolar transition frequency about its mean local environment gives
\begin{equation}
 \delta\omega_i
 =\frac{\partial\omega}{\partial V_{\alpha\beta}}\,
 \delta V_{\alpha\beta}({\bf R}_i)
 \equiv \frac{A_E q_i}{R_i^3},
 \label{eq:efg_shift}
\end{equation}
where $A_E$ includes all the constants, $q_i$ contains the angular factor, the sign/effective charge, and details of screening and local compensation.  Variations of the asymmetry parameter $\eta$ can be included in the same linear-response form and do not change the $1/R^3$ scaling.

\subsubsection{Random microstrains}
A charge neutral substitutional defect, vacancy, or interstitial appears at long distances as an elastic force-dipole $P_{\alpha\beta}$.  For
\begin{equation}
 H_{\rm el}=\frac12\int d^3r\,
 C_{\alpha\beta,\gamma\delta}u_{\alpha\beta}u_{\gamma\delta},
 \qquad
 u_{\alpha\beta}=\frac12(\partial_\alpha u_\beta+\partial_\beta u_\alpha),
 \label{eq:elastic_energy}
\end{equation}
the equation of mechanical equilibrium is the saddle-point equation,
\begin{equation}
 -\partial_\beta\!\left(C_{\alpha\beta,\gamma\delta}u_{\gamma\delta}\right)
 =f_\alpha({\bf r}),
 \qquad
 f_\alpha({\bf r})=-P_{\alpha\beta}\partial_\beta\delta({\bf r}-{\bf R}_i).
 \label{eq:elastic_EL}
\end{equation}
Expressing displacement $u_\alpha(\bf r)$ at Th nucleus in question in
terms of a force-dipole of the distant Th dopant and elastic Greens
function, $G_{\alpha\beta}$ (solution of \ref{eq:elastic_EL}) gives,
\begin{align}
 u_\alpha({\bf r})
 &=\int d^3r'\,G_{\alpha\beta}({\bf r}-{\bf r}')f_\beta({\bf r}')
 \nonumber\\
 &=P_{\beta\gamma}\partial_\gamma G_{\alpha\beta}({\bf r}-{\bf R}_i).
 \label{eq:elastic_displacement}
\end{align}
In three dimensions, the above Poisson-like equation gives
$G_{\alpha\beta}({\bf r})\sim1/r$ (up to the familiar tensor angular
structure), leading to scaling,
\begin{equation}
 u_\alpha({\bf r})\sim\frac{P}{C r^2},
 \qquad
 u_{\alpha\beta}({\bf r})\sim\frac{P}{C r^3}.
 \label{eq:strain_decay}
\end{equation}
A local strain changes bond lengths and angles around the observed Th ion and therefore changes its EFG.  To linear order,
\begin{equation}
 \delta\omega_i=\Lambda_{\alpha\beta}u_{\alpha\beta}({\bf R}_i)
 \equiv\frac{A_\epsilon q_i}{R_i^3}.
 \label{eq:strain_shift}
\end{equation}
A simple radial estimate $V_{zz}\propto r^{-3}$ gives $\delta V_{zz}/V_{zz}\simeq-3\epsilon$, and hence $A_\epsilon\sim3\omega_Q a^3$ up to geometry and elastic-response factors.  Unlike the direct EFG mechanism, this contribution is present even for electrically neutral defects.

For the line-shape derivation below, both mechanisms can therefore be
represented by the common model for the total line shift,
\begin{equation}
 S=\sum_i s_i,
 \qquad
 s_i=\frac{A q_i}{R_i^3},
 \label{eq:common_shift_model}
\end{equation}
with $q_i$ a zero-mean $O(1)$ random angular/sign factor.  We
take the simplest isotropic model $q_i=\pm1$ with equal probability,
observing that crystalline anisotropy only changes numerical
coefficients but not the central $1/r^3$ scaling behavior.

\subsection{Nearest-neighbor, single-impurity argument for the
  Lorentzian line shape}
\label{app:nearest_tail}
For a Poisson distribution of impurities of number density $n$, the
probability that the nearest impurity lies between $r$ and $r+dr$ is
the probability that the sphere of radius $r$ is empty times the
probability to find an impurity in the shell:
\begin{equation}
 P_r(r)dr
 =4\pi n r^2\exp\!\left(-\frac{4\pi n r^3}{3}\right)dr, 
 \label{eq:nearest_r_distribution}
\end{equation}
a standard Poisson result.  For the magnitude of the shift $s=A/r^3$,
\begin{equation}
 r=\left(\frac{A}{s}\right)^{1/3},
 \qquad
 \left|\frac{dr}{ds}\right|=\frac{A^{1/3}}{3s^{4/3}}.
 \label{eq:r_to_s}
\end{equation}
Therefore
\begin{equation}
 P_s(s)
 =P_r[r(s)]\left|\frac{dr}{ds}\right|
 =\frac{s_0}{s^2}\exp\!\left(-\frac{s_0}{s}\right),
 \qquad
 s_0\equiv\frac{4\pi}{3}nA.
 \label{eq:nearest_s_distribution}
\end{equation}
As pointed out by Stoneham \cite{Stoneham_1969}, at large shifts,
\begin{equation}
 P_s(s)\simeq\frac{s_0}{s^2},
 \qquad s\gg s_0,
 \label{eq:one_over_s2_tail}
\end{equation}
which is the characteristic tail of a Lorentzian distribution.  We
note that the divergence of the second moment
$\int ds\,s^2P_s(s)\to\infty$, is at the root of the breakdown of the
ordinary finite-variance central-limit theorem, which thereby does not
apply.

A real crystal supplies a short-distance cutoff $r\gtrsim a$, so the
continuum tail terminates at
\begin{equation}
 s_{\rm max}\sim\frac{A}{a^3}.
 \label{eq:lattice_cutoff}
\end{equation}
Thus the variance is not literally infinite on a lattice.  In the dilute limit $na^3\ll1$, however,
\begin{equation}
 \frac{s_{\rm max}}{s_0}\sim\frac{1}{na^3}\gg1,
 \label{eq:wide_lorentzian_window}
\end{equation}
so there is a parametrically broad regime over which the continuum $1/s^2$ tail and the Lorentzian description apply. 

\subsection{Detailed Poisson multi-impurity derivation of the
  Lorentzian line shape}
\label{app:detailed_lorentzian}
We have extended the above nearest-impurity tail-only analysis to a calculation of the full line-shape, where many uncorrelated impurities at positions $\{{\bf r}_i\}$ contribute to the inhomogeneous broadening.
The total random shift from $N$ impurities is then given by $\hat S =
\sum_i^N s_i$ with $s_i =\varepsilon q_i/r_i^3$. The corresponding
distribution of line shifts is given by
\begin{equation}
\begin{aligned}
\tilde P(S) &\equiv \int \frac{dk}{2\pi}\hat
P(k)e^{i k S}\\ & = \frac{1}{N!}\prod_i\left[\int n d^3r_i P({\bf
  r}_i)\right] \delta(S - \sum_i s_i) ,
  \end{aligned}
\label{PS}
\end{equation}
where the generating function is
\begin{eqnarray}
 \hat P(k)=\frac{1}{N!}\prod_i\bigg[\frac{1}{2}\sum_{q_i =\pm 1}\int n d^3r_i\ e^{-i k s_i}P(r_i)\bigg].
\end{eqnarray}

At this point, it is convenient to transition from this fixed-$N$ ensemble to the grand canonical analysis, with Poisson distributions of impurity occupations $\{n_r\}$ in each infinitesimal volume $dv_r$, with $N = \sum_r n_r \in \mathbb{Z}$ and $\sum_i s_i = \sum_r s_r\sum_a^{n_r}q_{r,a}$ ($n_r$ impurities with binary charges $q_{r,a}$). 
The generating function is then given by
\begin{eqnarray}
\hat P(k) &=&\prod_r\bigg[\sum_{n_r}\frac{1}{n_r!}(n dv_r)^{n_r}\nonumber\\
&&\times\frac{1}{2^{n_r}}\sum_{\{q_{r,a}=\pm 1\}} e^{-i k  s_r\sum_a^{n_r}q_{r,a} } e^{-n dv_r}\bigg],\nonumber\\
&=&e^{n\int d^3r(\cos k s_r-1)}
 =e^{n\tilde\varepsilon\int\frac{ds}{s^2}(\cos k s-1)}, \nonumber\\
 &=&e^{-\frac{1}{2}\pi n\tilde\varepsilon|k|},
\end{eqnarray}
where we used $\int_0^\infty dx\,(1-\cos x)/x^2=\pi/2$.
Fourier-transforming to $\tilde P(S)$ we then obtain the Lorentzian
%\begin{equation}
$\tilde P(S) = \frac{1}{\pi}\frac{\Gamma}{S^2 + \Gamma^2}$,
%\end{equation}
where the width is given by $\Gamma = \frac{1}{2}\pi n\tilde\varepsilon$.

To obtain the nuclear lineshape of thorium we then convolute its intrinsic line $I(\omega) = \gamma/[(\omega - \omega_0)^2 + \gamma^2]$ with $P(S)$, since disorder induces shifts $S$ in the line $\omega_0\rightarrow \omega_0 + S$. Since the intrinsic Th line has width $\gamma \ll \Gamma$ by $8$ orders of magnitude, $I(\omega)$ can be treated as a $\delta$-function, giving 
\begin{equation}
I_\text{inhom}(\omega) = \frac{1}{\pi}\frac{\Gamma}{(\omega-\omega_0)^2 + \Gamma^2}.
\end{equation}
This predicts the unusual Lorentzian (inhomogeneously broadened) line shape observed.

\section{\label{app:ab_initio_theory}\textit{Ab Initio} Calculations on Linewidth Broadening}
Atomistic calculations were performed using \textsc{lammps} \cite{Thompson2022LAMMPS,Plimpton1995LAMMPS}. For each defect density investigated, large periodic supercells (containing on average at least 10 dopants and of size up to 52$^3$ for the lowest densities) were constructed containing a dopant site at its center. Then, at a given density, taking a random distribution of dopants, each Ca ion has uniform probability of being replaced by a Th ion. This is repeated many times to obtain the statistical distribution of displacements containing, at least 200 Th probes at the lower densities. After such random doping of Th ions, for $\mathrm{C}_{3v}$ and $\mathrm{v}_{\mathrm{Ca}}$ configurations the local charge compensation geometry is then assigned randomly to each Th, selected from all symmetry equivalent configurations. 

The dimer has non-local charge compensation, modeled by randomly placed interstitial fluorine ions at least a distance of two conventional unit cells away from the dimer. In order to compute atomic displacements due to dopant-dopant interactions, an additional reference structure is built which consists of only a central dopant site.  The charge compensation structure of the defects is not yet firmly established, but this procedure is consistent with the non-local charge compensation assumption made in our DFT calculations that produced reasonably correct EFG values.

The structures were then all relaxed at fixed superlattice vectors, selected to minimize the energy of the reference structure, subject to the Buckingham pair potentials:
\[U_{ij} = A_{ij} e^{-r/\rho_{ij}} - \frac{C_{ij}}{r^6}\]
with $(A,\rho, C)$ equals $(3400 \text{~eV},0.2664 \text{~\AA},0)$ for Ca-F interactions, $(911.69 \text{~eV},0.2707 \text{~\AA}, 13.8 \text{~eV} \text{~\AA}^6)$ for F-F interactions, and $(3500 \text{~eV},0.2980 \text{~\AA},0)$ for Th-F interactions \cite{Dessovic2014ThCaF2}. These potential parameters were used in prior studies of Th:CaF$_2$ defect sites \cite{Dessovic2014ThCaF2}. The short-range potentials were cut off at $35$ \AA\ and the Coulomb interaction was evaluated using the particle-particle-particle-mesh solver implemented in \textsc{lammps}. The simulation parameters used were shown to converge atomic positions to a precision of $10^{-6}$ \AA. The structures were optimized utilizing the FIRE algorithm until the maximum interatomic force was below $10^{-8}$ eV\AA$^{-1}$ \cite{Bitzek2006FIRE}. By comparing each randomly placed Th in the optimized structures to the reference structures, the displacements of the bond vectors connecting each ion within one conventional unit cell of the dopant are computed (all ions within $a_0 = 5.55$\AA~ of the dopant). 

DFT calculations performed using the Vienna Ab-initio Simulation Package (VASP)~\cite{vasp1, vasp2, vasp3, vasp4, vasp5} in $2 \times 2 \times 2$ supercells of the conventional CaF$_2$ unit cell at the $\Gamma$-point provided the linear response function 
\[\Delta V_\mu = \sum_{n,\alpha} L_{\mu,n\alpha} u_{n\alpha}\]
where $u_{n\alpha}$ is the displacement of the bond connecting the Th ion to ion $n$ along the cartesian direction $\alpha$ and $\Delta V_\mu$ is the change of the $\mu = xx,yy,zz,xy,xz,yz$ component of the EFG, which maps the computed displacement fields onto EFGs. As a consistency check, multiple affine strains were applied to the DFT supercell and the EFGs at the Th ions were computed. The result was compared to the prediction made by the response function applied to the strains and was in good agreement, $\sim 4\%$. 
In order to best compare the effects of microstrain on linewidths across the various defect sites, the computed perturbations $\Delta V_\mu$ are added to the EFGs (experimentally extracted from the measured line centers), projected onto the principal axes of the EFG computed by DFT. 

%{\color{blue} EP: @LEO, please check and edit.\\
The computed EFGs were then projected through the spectrum of the NQR Hamiltonian, $H_1$, Eq.~(\ref{eq:qudrople_splitting}), and linewidths were extracted. In order to compare to experiment, these simulated linewidths were subject to the same analysis as in experiment. Fitting the linewidths to Eq.~(\ref{eq:FWHM}) in the main text yields, for the dimer, a ratio $\Gamma_{yy}/\Gamma_{zz} \sim 1.6$ consistently across simulated densities and $\Gamma_{yy}/\Gamma_{zz} \sim 0.6$ for $\mathrm{v}_{\mathrm{Ca}}$ and $\mathrm{C}_{3v}$ configurations. For the dimer configuration we find $\rho \sim -0.96$. The $\mathrm{v}_{\mathrm{Ca}}$ configuration fits with $\rho\sim 0$ and $\mathrm{C}_{3v}$ with $\rho \sim -0.8$. It should be noted that in the case of $\mathrm{C}_{3v}$ this coupling of EFG is not related, primarily, to correlation of displacements but rather that the EFG has $\eta = 0$ and this coupling is then primarily a property of the EFG symmetry. Directly probing the simulated distribution of EFGs as described above one obtains identical ratios. 

In contrast, when we set $\rho = 0$, the ratio of predictions from simulation for $\mathrm{v}_{\mathrm{Ca}}$ and $\mathrm{C}_{3v}$ do not change, but for the dimer the prediction changes to $\Gamma_{yy}/\Gamma_{zz} \sim 3.1$, differing substantially from the experimentally obtained EFG distributions. 

The origin of the linewidth hierarchy arises from the differing sensitivities of each transition to fluctuations in $V_{zz}$ or $V_{yy}$, as shown in Fig.~\ref{fig:EFG_spec}. In most cases, random displacements produce EFG distributions dominated by $V_{zz}$ broadening, causing a broad $b$ line. This can be understood at the level of the EFG produced by a classical point ion,
\[\bm V_{ii} = \frac{q_i}{4\pi\epsilon r^3}\left({3\bm n_i \bm n_i^T - 1}\right),\]
where the ion is at position $\bm r_i = r_i \bm n_i$. To this end, we consider a charge at $\bm r = (0,0,R)$ giving
\[(V_{xx},V_{yy},V_{zz}) = \frac{q}{4\pi\epsilon R^3}(-1,-1,2).\]
A small displacement along the $z$ axis then produces $\Gamma_{yy}/\Gamma_{zz} = 1/2$, which is near the ratio extracted from the simulations for the $\mathrm{C}_{3v}$ and $\mathrm{v}_{\mathrm{Ca}}$ sites. For any number of independent axial displacements this ratio will remain exactly $1/2$, and the only way for this to change is by some form of correlated displacements. Considering uniform dilation of the lattice constant one obtains $\Gamma_{yy}/\Gamma_{zz} = \frac{1+\eta}{2} < 1$.
Thus, ordinarily $V_{zz}$ dominates the fluctuations and correlated or non-affine displacements are required to produce the linewidth hierarchy observed in experiment. This is the expected pattern when the microstrain produces lattice constant variations. The fact that the dimer produces a ratio closer to two is a consequence of correlated displacements (because the defect site is charged, ions of like charge move similarly), primarily involving the first fluorine shell surrounding the dopant. Because such correlated movement is required, it is not evident in the present simulations whether this result will be preserved under stronger charge screening from a quasi-local defect complex.

\nocite{*}

\bibliography{my_bib}% Produces the bibliography via BibTeX.

\end{document}